\documentclass[10pt,journal,twocolumn,twoside]{IEEEtran}
\usepackage{setspace}
\usepackage{cite}
\usepackage{color}
\usepackage{amsmath}
\usepackage{amsfonts}
 \usepackage{amssymb}
\usepackage{multicol} 
\usepackage{graphicx}
\usepackage{algorithm}
\usepackage{algorithmic}
\usepackage{cleveref}
\usepackage{epsfig}
 \usepackage{epstopdf}
\usepackage{url}
 \usepackage[noblocks]{authblk}
 \usepackage{makecell}
 \usepackage{verbatim}
\begin{document}

\title{Joint Port Selection and Beamforming Design for Fluid Antenna Assisted Integrated Data and Energy Transfer}
\author{Long Zhang, Halvin Yang,~\IEEEmembership{Student Member,~IEEE}, Yizhe Zhao,~\IEEEmembership{Member,~IEEE},

Jie Hu,~\IEEEmembership{Senior Member,~IEEE},

\thanks{Long Zhang, Yizhe Zhao and Jie Hu are with the School of Information and Communication Engineering, University of Electronic Science and Technology of China, Chengdu 611731, China (e-mail: l.zhang@std.uestc.edu.cn;yzzhao@uestc.edu.cn; hujie@uestc.edu.cn).

Halvin Yang are with the Department of Electronic and Electrical Engineering, University College London, WC1E 7JE London, U.K.(e-mail: uceehhy@ucl.ac.uk).}}

\maketitle


\begin{abstract}
Integrated data and energy transfer (IDET) has been of fundamental importance for providing both wireless data transfer (WDT) and wireless energy transfer (WET) services towards low-power devices. Fluid antenna (FA) is capable of exploiting the huge spatial diversity of the wireless channel to enhance the receive signal strength, which is more suitable for the tiny-size low-power devices having the IDET requirements. In this letter, a multiuser FA assisted IDET system is studied and the weighted energy harvesting power at energy receivers (ERs) is maximized by jointly optimizing the port selection and transmit beamforming design under imperfect channel state information (CSI), while the signal-to-interference-plus-noise ratio (SINR) constraint for each data receiver (DR) is satisfied. An efficient algorithm is proposed to obtain the suboptimal solutions for the non-convex problem. Simulation results evaluate the performance of the FA-IDET system, while also demonstrate that FA outperforms the multi-input-multi-output (MIMO) counterpart in terms of the IDET performance, as long as the port number is large enough.

\end{abstract}


%
\IEEEpeerreviewmaketitle

\section{Introduction}
In the era of 6G and Internet of Things (IoT), numerous low-power devices are swarming into wireless networks, which imposes great challenges on energy supplement for these devices \cite{9349624}. Integrated data and energy transfer (IDET) technique is capable of enabling both wireless data transfer (WDT) and wireless energy transfer (WET) towards the low-power devices, which can effectively solve the energy supply problem and prolong their service life. However, due to the limited space, most of low-power devices could only be equipped with single antenna, which causes low energy efficiency.

In recent years, fluid antenna (FA)\cite{10146274,9264694,9715064,9650760} and movable antenna \cite{10318061,10243545} are considered as new prospects to boom 6G and have aroused great concerns due to their tremendous flexibility and reconfigurability. Both of them could adjust their receiving antenna to the desired position in a given region, in order to achieve an additional receive gain from the various channel fading in the space.  However, movable antennas are more suitable for the transmitter such as the base station, due to the higher energy consumption and the larger required physical space. FA can be switched flexibly to one of $N$ fixed locations (defined as ports) in a linear space, in which the optimal location is usually determined according to users' requirement, such as maximizing the signal-to-interference-plus-noise ratio (SINR). Although traditional multiple-input multiple-output (MIMO) technique provides huge multiplexing gain for receivers, it is difficult to deploy multiple antennas at low-power devices. FA is capable of exploiting high spatial diversity to improve energy harvesting efficiency, which is more suitable for low-power devices.

FA was firstly proposed by Wong \textit{et al.} in \cite{9264694}, in which the outage probability was derived and it was demonstrated that the FA system can outperform $L$-antenna maximum ratio combining (MRC) system when the port number $N$ is large enough. In order to tackle with the problem of redundant channel estimation, machine learning was considered to efficiently select ports from a few observed ones \cite{9715064}. In addition, Wong \textit{et al.} \cite{9650760} studied fluid antenna multiple access (FAMA) to achieve a higher multiplex gain in the multi-user system. Further, a specific version, namely slow fluid antenna multiple access (s-FAMA), was then proposed in \cite{10066316}, where the optimal port was selected in the space where the fading envelopes of the other users were in a deep fade.

However, none of existing works studied FA assisted IDET. Thanks to the tremendous space diversity of FA system, the IDET performance could be further improved compared to traditional fixed-position antennas. Our contributions are then summarised as follows:
\begin{itemize}
\item We originally study a FA assisted IDET system, which consists of one transmitter and multiple data receivers (DRs) and energy receivers (ERs). The DRs and the ERs are all equipped with a single fluid antenna, which is able to dynamically adjust the port selection according to the wireless channels.
\item The port selection and beamforming design are jointly optimized under imperfect channel state information (CSI), in order to maximize the weighted energy harvesting power at ERs by guaranteeing the SINR constraints of each DR. A semidefinite relaxation (SDR) based alternating optimization (AO) approach is proposed to obtain the sub-optimal solutions for the non-convex problem. A low-complexity method is also discussed to reduce the overhead of CSI acquisition and complexity of the proposed algorithm.
\item The performance of the FA assisted IDET system is evaluated and compared with traditional MIMO by simulation results, which also provides some novel insights for the system design.
\end{itemize}

\textit{Notations:} In this paper, $\mathbb{C}$ denotes the set of complex numbers. $|x|$ represents the absolute value of a complex number $x$, while $||\boldsymbol{x}||$ represents the 2-norm of the vector $\boldsymbol{x}$. Moreover, $(\cdot)^H$ denotes the conjugate operation of complex vector or matrix

\section{System Model}\label{section:architecture}
\begin{figure}
	\centering
	\includegraphics[width=0.7\linewidth]{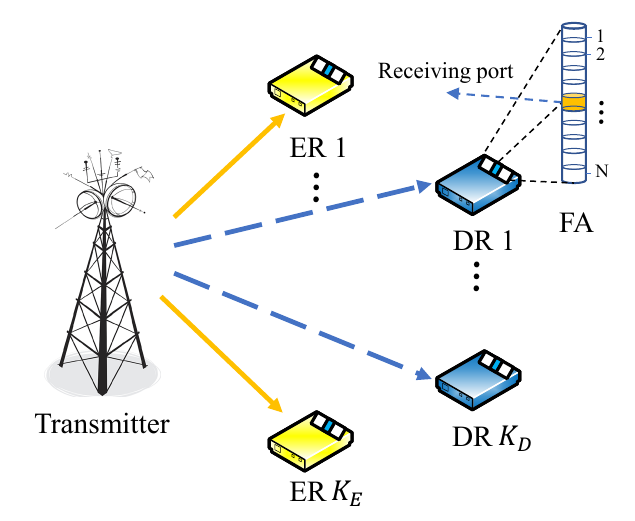}
	\caption{System model.}
	\label{fig:system_model}
\end{figure}

As shown in Fig. \ref{fig:system_model}, a FA assisted IDET system is studied, which consists of a transmitter, $K_D$ DRs and $K_E$ ERs. The transmitter is equipped with $M$ fixed-position antennas which are spatially far apart so that the wireless channel of each antenna is independent. Each DR or ER is equipped with one fluid antenna. The length of fluid antenna at DR $i$ (or ER $j$) is denoted as $W_i\lambda$ (or $W_j\lambda$), where $\lambda$ is the wavelength of the radio frequency (RF) signal \cite{9650760}.The set of DRs and ERs are denoted by $\mathcal{K}_D=\{1,2,\cdots, {K_D}\} $ and $\mathcal{K}_E=\{1,2,\cdots, K_E\}$, respectively. The port number of DR $i$ and ER $j$ are denoted $N_i^D$ and $N_j^E$ respectively. For the simplicity, we assume all the DRs and ERs have the same port number as $N_i^D=N_j^E=N$.
 \raggedbottom
\subsection{Wireless channel model}
The quasi-static flat fading channels are conceived between the transmitter and receivers, while the CSI of all links are assumed to be imperfect. Denote the channel between the transmitter and DR $i$ and that between the transmitter and ER $j$ as $\mathbf{H}_{i}\in\mathbb{C}^{M\times N}$ and $\mathbf{G}_{j}\in\mathbb{C}^{M\times N}$, respectively. Taking into account the channel estimation error, the ideal channel $\mathbf{H}_{i}$ and $\mathbf{G}_{j}$ are given by
\begin{align}
\mathbf{H}_{i}=\rho_i \hat {\mathbf{H}}_{i}+\sqrt{1-\rho_i^2} \Delta\mathbf{H}_{i}, \notag\\
\mathbf{G}_{j}=\rho_j \hat {\mathbf{G}}_{j}+\sqrt{1-\rho_j^2} \Delta\mathbf{G}_{j},
\end{align}
where $\rho_i$ and $\rho_j$ are the channel estimation accuracy parameter, $\hat {\mathbf{H}}_{i}$ and $\hat {\mathbf{G}}_{j}$ represent the estimated CSI from the transmitter to DR $i$ and ER $j$, respectively. $\Delta\mathbf{H}_{i}$ and $\Delta\mathbf{G}_{j}$ denote the channel estimate error, whose elements are Gaussian complex variables with zero mean and variance of $\sigma_h^2$ and $\sigma_g^2$, which are equal to the large scale path-loss of each link. Since the ports in the fluid antenna are close enough, the wireless channels of different ports are correlated with each other, while the correlation is characterized as a parameter $\mu_i$ \cite{10066316}. The $m$-th row and $n$-th column of $\hat {\mathbf{H}}_{i}$ or $\hat {\mathbf{G}}_{j}$ denotes the wireless channel between the $m$-th transmitter antenna and the $n$-th port of DR $i$ or ER $j$, which is expressed as
\begin{align}\label{channel}
\hat{{h}}_{i,m,n}(\textrm{or}\ \hat{{g}}_{j,m,n})=&(\sqrt{1-\mu_i^2}{x}_{i,m,n}+\mu_i {x}_{i,m,0})\notag\\
+&j(\sqrt{1-\mu_i^2}{y}_{i,m,n}+\mu_i {y}_{i,m,0}),
\end{align}
where ${x}_{i,m,0},\cdots,{x}_{i,m,N}$ and ${y}_{i,m,0},\cdots,{y}_{i,m,N}$ are all independently Gaussian distributed having the zero mean and the variance of 0.5. According to \cite{Wong_EL}, the correlation parameter $\mu_i$ is given by
\begin{align}
\mu_i=\sqrt{2}\sqrt{_{1}F_{2}(\frac{1}{2};1;\frac{3}{2};-\pi^2W_i^2)-\frac{J_1(2\pi W_i)}{2\pi W_i}},
\end{align}
where $_{1}F_{2}(\cdot;\cdot;\cdot)$ is the generalized hypergeometric function and $J_1(\cdot)$ is the first-order Bessel function of the first kind.

 \raggedbottom
\subsection{Signal model}
In order to simultaneously transmit signals towards all the DRs and ERs, beamforming is required at the transmitter, while each DR and ER is allocated with one beamforming vector. Then, the transmitted signal for transmitter is expressed as
\begin{align}\label{tx signal}
\boldsymbol{x}=\sum_{i\in\mathcal{K}_D}\boldsymbol{w}_is_{D,i}+\sum_{j\in\mathcal{K}_E}\boldsymbol{v}_js_{E,j},
\end{align}
where $\boldsymbol{w}_i\in\mathbb{C}^{M\times1}$ and $\boldsymbol{v}_j\in\mathbb{C}^{M\times1}$ are the beamforming vectors allocated for DR $i$ and ER $j$, while $s_{D,i}$ and $s_{E,j}$ are the Gaussian distributed information and energy signals having the unit power, respectively. Under a given transmitting power $P$, we have $\mathbb{E}(\boldsymbol{x}^H\boldsymbol{x})=\sum_{i\in\mathcal{K}_D}||\boldsymbol{w}_i||^2+\sum_{j\in\mathcal{K}_E}||\boldsymbol{v}_j||^2\leq P$.

Denote $ \mathbf{r}_{D,i,n}$ and $ \mathbf{r}_{E,j,n}$ as the  port activation indicator of DR $i$ and ER $j$, respectively. Since there is only one port activated for receiving signals, only the $n$-th element of $ \mathbf{r}_{D,i,n}$ or $ \mathbf{r}_{E,j,n}$ is 1 and the rest is 0. Therefore, the received signal at the $n$-th port of DR $i$ is expressed as
\begin{align}
y_{i,n}^{DR}=\mathbf{r}_{D,i,n}^{H}\mathbf{H}_{i}^{H}\boldsymbol{x}+z_{i,n},
\end{align}
\begin{figure*}[ht]
\centering
\hrulefill
\vspace*{8pt}
\begin{equation}
\label{SINR}
\mathrm{SINR}_{i,n}=\frac{\rho_i^2|\mathbf{r}_{D,i,n}^{H}\hat{\mathbf{H}}_i^{H}\boldsymbol{w}_i|^2}{\sum\limits_{{k}\in\mathcal{K}_D,\atop k\neq i}{\rho_i^2|\mathbf{r}_{D,i,n}^{H}\hat{\mathbf{H}}_i^{H}\boldsymbol{w}_{k}|^2}+\sum\limits_{ {j}\in\mathcal{K}_E}{\rho_i^2|\mathbf{r}_{D,i,n}^{H}\hat{\mathbf{H}}_i^{H}\boldsymbol{v}_j|^2}+(1-\rho_i^2)\sigma_h^2(\sum\limits_{i\in\mathcal{K}_D}||\boldsymbol{w}_i||^2+\sum\limits_{j\in\mathcal{K}_E}||\boldsymbol{v}_j||^2)+\sigma_i^2}.
\end{equation}
\end{figure*}
where $z_{i,n}$ is the additive white Gaussian noise (AWGN) having the zero mean and the variance of $\sigma_{i}^2$ at the $n$-th port of DR $i$. It is assumed that the DR cannot eliminate the interference arisen from energy signals, thus the receive SINR of DR $i$ at $n$-th port can be expressed as Eq. \eqref{SINR}.

On the other hand, by ignoring the noise at ER antenna, the energy harvesting power of ER $j$ at $n$-th port is expressed by
\begin{align}
\mathrm{E}_{j,n}&=\sum_{i\in\mathcal{K}_D}{\rho_j^2|\mathbf{r}_{E,j,n}^{H}\hat{\mathbf{G}}_j^{H}\boldsymbol{w}_i|^2}+\sum_{{k}\in\mathcal{K}_E}{\rho_j^2|\mathbf{r}_{E,j,n}^{H}\hat{\mathbf{G}}_j^{H}\boldsymbol{v}_{k}|^2}\notag\\
&+(1-\rho_j^2)\sigma_g^2(\sum\limits_{i\in\mathcal{K}_D}||\boldsymbol{w}_i||^2+\sum\limits_{k\in\mathcal{K}_E}||\boldsymbol{v}_k||^2).
\end{align}

\section{Joint Port Selection and Beamforming Design}
\subsection{Problem Formulation}
In this paper, we aim to jointly optimize the port selection at the DRs and ERs as well as the beamforming design at the transmitter, in order to maximize the weighted energy harvesting power at ERs under the constraints of SINR at each DR. The weighted energy harvesting power at ERs is expressed as

\begin{align}
\mathrm{E}_{H}=\sum_{j\in\mathcal{K}_E}{\beta_{j}\mathrm{E}_{j,n}}=\sum_{i\in\mathcal{K}_D}{\boldsymbol{w}_i^{H}\boldsymbol{S}_{r_E}\boldsymbol{w}_i}+\sum_{j\in\mathcal{K}_E}{\boldsymbol{v}_j^{H}\boldsymbol{S}_{r_E}\boldsymbol{v}_j},
\end{align}
where $\boldsymbol{S}_{r_E}=\sum_{j\in\mathcal{K}_E}\beta_{j}(\rho_j^2\hat{\mathbf{G}}_j\mathbf{r}_{E,j,n}\mathbf{r}_{E,j,n}^{H}\hat{\mathbf{G}}_j^{H}+(1-\rho_j^2)\sigma_g^2\mathbf{I})$, $\beta_{j}\geq0$ denotes the energy weight for ER $j$, while a larger value indicates a higher energy harvesting requirement of ER $j$ compared to other ERs \cite{8941080}. Denoted the port activation indicator set of all receivers as $\mathbf{r}=\{\mathbf{r}_D,\mathbf{r}_E\}$, where $\mathbf{r}_D=\{\mathbf{r}_{D,i,n},i=1,\cdots,K_D\}$ and $\mathbf{r}_E=\{\mathbf{r}_{E,j,n},j=1,\cdots,K_E\}$, the optimization problem can be formulated as
\begin{align}
\text{(P1)}:\max_{\left\{\boldsymbol{w}_i\right\},\left\{\boldsymbol{v}_j\right\},\mathbf{r}} \quad & \mathrm{E}_H, \label{obj1}\\
s.t. \ \ \ \ \quad &\mathrm{SINR}_{i,n}\geq\gamma_{i},\forall i\in\mathcal{K}_D, \tag{\ref{obj1}{a}}\label{obj1a}\\
&\sum_{i\in\mathcal{K}_D}||\boldsymbol{w}_i||^2+\sum_{j\in\mathcal{K}_E}||\boldsymbol{v}_j||^2\leq P , \tag{\ref{obj1}{b}}\label{obj1b}\\
&\sum_{n=1}^{N} \mathbf{r}_{D,i,n}=1,\forall i\in\mathcal{K}_D,\notag\\
&\mathbf{r}_{D,i,n}=0 \ \mathrm{or}\ 1, n=1,\cdots,N,\tag{\ref{obj1}{c}}\label{obj1c}\\
&\sum_{n=1}^{N} \mathbf{r}_{E,j,n}=1,\forall j\in\mathcal{K}_E,\notag\\
&\mathbf{r}_{E,j,n}=0 \ \mathrm{or}\ 1, n=1,\cdots,N,\tag{\ref{obj1}{d}}\label{obj1d}
\end{align}
 where \eqref{obj1a} represents the SINR constraints of each DR, \eqref{obj1b} represents the limited transmitted signal power. Note that the beamforming vectors and the port activation indicators are intricately coupled in both the objective function and SINR constraints, which is a non-convex optimization problem and hard to obtain the global solutions. Therefore, the AO algorithm is applied for obtaining the sub-optimal solutions.
 \raggedbottom
\subsection{FA assisted IDET}
Firstly, we fix the port activation indicators $\mathbf{r}$ to optimize the information beamforming vectors \{$\boldsymbol{w}_i$\} and the energy beamforming vectors $\{\boldsymbol{v}_j\}$. Denote $\boldsymbol{W}_i=\boldsymbol{w}_i\boldsymbol{w}_i^H,\forall i\in\mathcal{K}_D$ and $\boldsymbol{V}_E=\sum_{j\in \mathcal{K}_E}\boldsymbol{v}_j\boldsymbol{v}_j^H$, we have $\text{rank}(\boldsymbol{W}_i)\leq1,\forall i\in\mathcal{K}_D$ and $\text{rank}(\boldsymbol{V}_E)\leq \text{min}\left\{M,K_E\right\}$. By ignoring the rank constraints on $\boldsymbol{W}_i$ and $\boldsymbol{V}_E$, the SDR of ($\text{P1}$) can be formulated as
\begin{align}
\text{(P2)}:
\max_{\left\{\boldsymbol{W}_i\right\},\boldsymbol{V}_E} \ & \sum_{i\in\mathcal{K}_D}{\text{tr}(\boldsymbol{S}_{r_E}\boldsymbol{W}_i})+\text{tr}(\boldsymbol{S}_{r_E}\boldsymbol{V}_E) \label{obj3}\\
s.t. \ \ \ \ \ &\frac{\text{tr}(\boldsymbol{F}_{r_D}\boldsymbol{W}_i)}{\gamma_i}-\sum_{{k}\in\mathcal{K}_D,\atop k\neq i}\text{tr}(\boldsymbol{D}_{r_D}\boldsymbol{W}_{k})\notag\\
&-\text{tr}(\boldsymbol{D}_{r_D}\boldsymbol{V}_E)-\sigma_i^2\geq0,\forall i\in\mathcal{K}_D\tag{\ref{obj3}{a}}\label{obj3a} \\
&\sum_{i\in\mathcal{K}_D}\text{tr}(\boldsymbol{W}_i)+\text{tr}(\boldsymbol{V}_E) \leq P \tag{\ref{obj3}{b}}\label{obj3b}\\
&\boldsymbol{W}_i\geq 0,\forall i\in\mathcal{K}_D, \ \boldsymbol{V}_E\geq 0 \tag{\ref{obj3}{c}}\label{obj3c}
\end{align}
where $\boldsymbol{F}_{r_D}=\rho_i^2\hat{\mathbf{H}}_i\mathbf{r}_{D,i,n}\mathbf{r}_{D,i,n}^H\hat{\mathbf{H}}_i^H-\gamma_i(1-\rho_i^2)\sigma_h^2\mathbf{I}$ and $\boldsymbol{D}_{r_D}=\rho_i^2\hat{\mathbf{H}}_i\mathbf{r}_{D,i,n}\mathbf{r}_{D,i,n}^H\hat{\mathbf{H}}_i^H+(1-\rho_i^2)\sigma_h^2\mathbf{I}$. According to the rank reduction theorem in \cite{huang2009rank}, the optimal solutions for any solvable separable semidefinite program (SDP) satisfy $\sum_{i\in\mathcal{K}_D}\text{rank}(\boldsymbol{W}_i^{*})^2+\text{rank}(\boldsymbol{V}_E^{*})^2\leq K_D+1$. Meanwhile, the SINR constraints indicate that the solutions always satisfy $\boldsymbol{W}_i^{*}\neq\boldsymbol{0}$ and $\text{rank}(\boldsymbol{W}_i^{*})\geq 1,\forall i\in\mathcal{K}_D$. Thus, it is obtained that the optimal solutions of (P2) satisfy $\text{rank}(\boldsymbol{W}_i^{*}) = 1,\forall i\in\mathcal{K}_D$ and $\text{rank}(\boldsymbol{V}_{E}^{*}) \leq 1$. Since the rank constraints of original problem are satisfied, (P2) has the same optimal solutions with (P1) when $\mathbf{r}$ is fixed. Note that when $\mathbf{r}$ is fixed, (P2) is a SDP problem, which is convex and can be efficiently solved by standard solvers. Then the optimal beamforming vectors $\boldsymbol{w}_i^{*}$ and $\boldsymbol{v}_j^{*}$ can be recovered through eigenvalue decomposition (EVD) over the obtained $\boldsymbol{W}_i^{*}$ and $\boldsymbol{V}_E^{*}$.

Next, we fix the optimal beamforming vectors to optimize the port activation indictors $\mathbf{r}$. When  the optimal beamforming vectors are fixed, (P1) can be reformulated as
\begin{align}
\text{(P3)}:\notag\\
\ \max_{\{\mathbf{R}_{E,j,n}\},\{\mathbf{R}_{D,i,n}\}}\ & \sum_{j\in\mathcal{K}_E}\text{tr}(\hat{\mathbf{G}}_{j}^H\Omega\hat{\mathbf{G}}_{j}\mathbf{R}_{E,j,n}) \label{obj4}\\
s.t. \ \quad & \frac{\text{tr}(\hat{\mathbf{H}}_{i}^{H}\boldsymbol{W}_{i}^{*}\hat{\mathbf{H}}_{i}\mathbf{R}_{D,i,n})}{\text{tr}(\hat{\mathbf{H}}_{i}^{H}\boldsymbol{\Psi}\hat{\mathbf{H}}_{i}\mathbf{R}_{D,i,n})+Z}\geq \gamma_i\tag{\ref{obj4}{a}}\label{obj4a}\\
&\sum_{n=1}^{N}\mathrm{diag}(\mathbf{R}_{E,j,n})=1,j=1,\cdots,K_E\tag{\ref{obj4}{b}}\label{obj4b}\\
&\sum_{n=1}^{N}\mathrm{diag}(\mathbf{R}_{D,i,n})=1,i=1,\cdots,K_D\tag{\ref{obj4}{c}}\label{obj4c}\\
&\mathrm{diag}(\mathbf{R}_{E,j,n})=1 \ \mathrm{or} \ 0, n=1,\cdots,N\tag{\ref{obj4}{d}}\label{obj4d}\\
&\mathrm{diag}(\mathbf{R}_{D,i,n})=1 \ \mathrm{or} \ 0, n=1,\cdots,N\tag{\ref{obj4}{e}}\label{obj4e}
\end{align}
where $\mathbf{R}_{E,j,n}=\mathbf{r}_{E,j,n}\mathbf{r}_{E,j,n}^H$, $\mathbf{R}_{D,i,n}=\mathbf{r}_{D,i,n}\mathbf{r}_{D,i,n}^H$, $\Omega=\sum_{i\in\mathcal{K}_D}\boldsymbol{W}_i^{*}+\boldsymbol{V}_E^{*}$, $\boldsymbol{\Psi}=\sum_{k\in\mathcal{K}_D, k\neq i}\boldsymbol{W}_k^{*}+\boldsymbol{V}_E^{*}$, $Z=(1-\rho_i^2)\sigma_h^2(\sum_{i\in\mathcal{K}_D}\text{tr}(\boldsymbol{W}_i^{*})+\text{tr}(\boldsymbol{V}_E^{*}))/\rho_i^2+\sigma_i^2/\rho_i^2$. As the port selection of all DRs and ERs are independent of each other, their primary objective is to select the most desirable port at their respective locations. Since the energy port activation indicators only exist in the objective function, we aim to maximize the energy harvesting power at each ER to obtain the indicator, which can be formulated as
\begin{align}
\text{(P4)}:\ \max_{\mathbf{R}_{E,j,n}}\ \text{tr}(\hat{\mathbf{G}}_{j}^H\Omega\hat{\mathbf{G}}_{j}\mathbf{R}_{E,j,n})
\end{align}
 Since $\mathbf{R}_{E,j,n}$ is a diagonal matrix with 1 for the $n$-th diagonal element and 0 for the rest, the index of optimal energy port can be easily obtained by
\begin{align}
n_E^*=\mathrm{argmax} \ \mathrm{diag} (\hat{\mathbf{G}}_{j}^H\Omega\hat{\mathbf{G}}_{j})_n,n=1,\cdots,N,
\end{align}
where $\mathrm{diag}(\cdot)_n$ denotes the operation of taking the $n$-th diagonal element. On the other hand, the data port activation indicator can be obtained by maximizing the SINR at each DR, which can be formulated as
\begin{align}
\text{P5}:\ \max_{\mathbf{R}_{D,i,n}}\ \frac{\text{tr}(\hat{\mathbf{H}}_{i}^{H}\boldsymbol{W}_{i}^{*}\hat{\mathbf{H}}_{i}\mathbf{R}_{D,i,n})}{\text{tr}(\hat{\mathbf{H}}_{i}^{H}\boldsymbol{\Psi}\hat{\mathbf{H}}_{i}\mathbf{R}_{D,i,n})+Z}
\end{align}
 Thus the index of optimal data port is obtained by
\begin{align}
n_D^*=\mathrm{argmax} \ \frac{\mathrm{diag} (\hat{\mathbf{H}}_{i}^{H}\boldsymbol{W}_{i}^{*}\hat{\mathbf{H}}_{i})_n}{\mathrm{diag}(\hat{\mathbf{H}}_{i}^{H}\boldsymbol{\Psi}\hat{\mathbf{H}}_{i})_n+Z},n=1,\cdots,N.
\end{align}
Since the objective value of (P1) is non-decreasing by alternately optimizing the beamforming vectors and the port selection indexes, there always exists an upper bond of (P1), which indicates that the proposed SDR based AO algorithm can converge to a sub-optimal solution. Given a solution accuracy $\epsilon$, the computational complexity of solving (P2) in each iteration is $\mathcal{O}((K_D^3M^{3.5}+K_D^4)\text{log}(1/\epsilon))$ \cite{6860253}, while that of solving (P3) is $\mathcal{O}(N(K_D+K_E))$. Thus the total computational complexity of the proposed algorithm is $\mathcal{O}(t[(K_D^3M^{3.5}+K_D^4)\text{log}(1/\epsilon)+N(K_D+K_E)])$, where $t$ is the iteration number. The details of the proposed SDR based AO algorithm are summarized in Algorithm \ref{power}.
\begin{algorithm}[t]
    \renewcommand{\algorithmicrequire}{\textbf{Input:}}
	\renewcommand{\algorithmicensure}{\textbf{Output:}}
	\caption{Proposed Alternating Optimization Algorithm}
    \label{power}
    \begin{algorithmic}[1]
        \REQUIRE  Initialize the port activation indicator set $\mathbf{r}=\mathbf{r}^{(0)}$, set iteration number $t=0$ and give the maximum iteration number $T_{max}$ and tolerance value $\tau$ ;
	    \ENSURE the optimal solution $\{\boldsymbol{w}_i^{*}\}$, $\boldsymbol{v}_j^{*}$ and $\mathbf{r}^*$;
        \STATE \textbf{repeat};
        \STATE \quad  Obtain $\{\boldsymbol{W}_i^{(t+1)}\}$ and $\boldsymbol{V}_E^{(t+1)}$ by solving the problem (P2) for a given $\mathbf{r}^{(t)}$;
        \STATE \quad  Given $\{\boldsymbol{W}_i^{(t+1)}\}$ and $\boldsymbol{V}_E^{(t+1)}$, obtain $\mathbf{r}^{(t+1)}$ according to (13) and (15), respectively.
        \STATE \textbf{until} $t\leq T_{max}$ or $\mathrm{E}_H^{(t+1)}-\mathrm{E}_H^{(t)}\leq\tau$;
        \STATE  Perform EVD for $\{\boldsymbol{W}_i^{(t+1)}\}$ and $\boldsymbol{V}_E^{(t+1)}$ and then obtain the optimal beamforming vectors $\{\boldsymbol{w}_i^{*}\}$ and $\boldsymbol{v}_j^{*}$.
        \STATE \textbf{return} $\{\boldsymbol{w}_i^{*}\}$, $\boldsymbol{v}_j^{*}$ and $\mathbf{r}^*=\mathbf{r}^{(t+1)}$.
    \end{algorithmic}
\end{algorithm}
\subsection{Low-complexity method}
In this subsection, we propose a low-complexity method to reduce the overhead of CSI acquisition. Since it is infeasible to acquire the CSI of all ports, we only estimate the channel of a subset of 'selected' ports for each FA \cite{9992289}. Thanks to the strong spatial correlation between ports, the channels are very similar between adjacent ports and thus we can estimate the channel at intervals of several ports. Initially, the transmitter broadcasts the pilot-training symbols to the reference (i.e., first) port (i.e., 'selected' port) and the CSI between the transmitter and the reference port is obtained. Then, assuming that the subsequent $L-1$ ports' CSI are similar to the previous estimated port due to the strong spatial correlation, we skip the CSI acquisition process of these ports and estimate the CSI between the transmitter and the $(L+1)$-th port. Repeat the above process until the CSI of all 'selected' ports are estimated. Compared with the conventional channel estimation for all the $N$ ports, our proposed low-complexity method only need to acquire the CSI of $K=\lceil\frac{N}{L}\rceil$ ports, which reduces both the overhead of CSI acquisition and the complexity of the proposed algorithm.

\section{Simulation Results}\label{section:sim}
In this section, the performance of FA assisted IDET system is evaluated by simulation and compared with traditional MIMO assisted IDET by setting the same antenna size. Without specific statement, the number of transmitter antennas is set to $M=4$. The size of all FA is set the same as $W$ and the channel estimation accuracy parameter between the transmitter and each receiver is set the same as $\rho$. The AWGN power at the receive antenna is set to $\sigma_i^2=-50$ dBm while the transmitted power is set to 30 dBm. The distance between the transmitter and all the ERs are set to 3 m, while that between the transmitter and all the DRs are set to 10 m. The pathloss factor is set to 2.7. It is assumed that there are 2 ERs and 2 DRs in the system. In addition, the energy weight $\beta_j$ for all the ERs are set the same as $\beta_j=1, \forall j\in\mathcal{K}_E$.
\begin{figure}[t]
	\centering
	\includegraphics[width=0.8\linewidth]{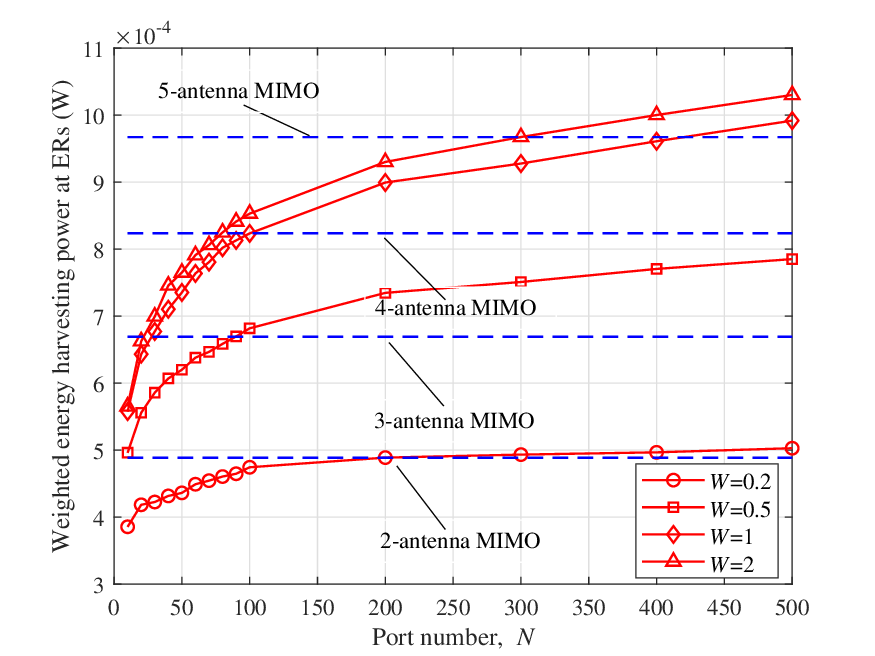}
	\caption{Weighted energy harvesting power at ERs versus the port number $N$ with different antenna size $W$.}
	\label{fig:IDET_2}
\end{figure}

Fig. \ref{fig:IDET_2} depicts the weighted energy harvesting power at ERs versus the port number $N$ with different antenna size $W$. In the MIMO assisted IDET benchmark, the transmitting beamforming vectors for DRs and ERs as well as the receiving beamforming vectors at DRs are jointly optimized to maximize the weighted energy harvesting power at ERs under the SINR constraints of DRs. The SINR threshold is set to 10 dB. In order to evaluate the best performance of FA assisted IDET system, both $\rho$ and $L$ are set to 1. Observe from Fig. \ref{fig:IDET_2} that port number can significantly strengthen the energy harvesting power at ERs. Moreover, a larger $W$ achieves the higher energy harvesting power, since the correlation among ports decreases with $W$, which results in a better IDET performance. Compared with MIMO assisted IDET with the same antenna size, FA always achieves a better WET performance as long as the number of ports is large enough.

\begin{figure}[t]
	\centering
	\includegraphics[width=0.8\linewidth]{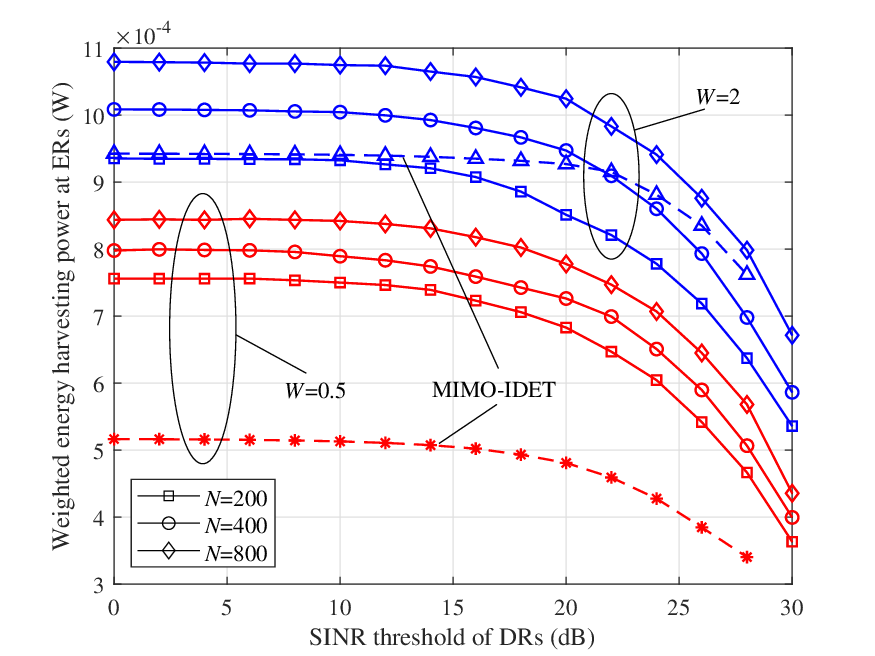}
	\caption{Weighted energy harvesting power at ERs versus the SINR threshold with different antenna size $W$ and port number $N$.}
	\label{fig:IDET_4}
\end{figure}
Fig. \ref{fig:IDET_4} depicts the weighted energy harvesting power at ERs versus the SINR threshold, where various antenna size $W$ and different port number $N$ are conceived. Observe from Fig. \ref{fig:IDET_4} that when the number of ports is larger, ERs harvest more energy. Moreover, a larger antenna size $W$ results in a better IDET performance since the channel correlation between different ports decreases with $W$ and it is more likely to select an optimal port having the peak envelop of the wireless channel. Compared with MIMO counterpart, FA assisted IDET system achieves better energy harvesting performance when the port number is large enough.

\begin{figure}[t]
	\centering
	\includegraphics[width=0.8\linewidth]{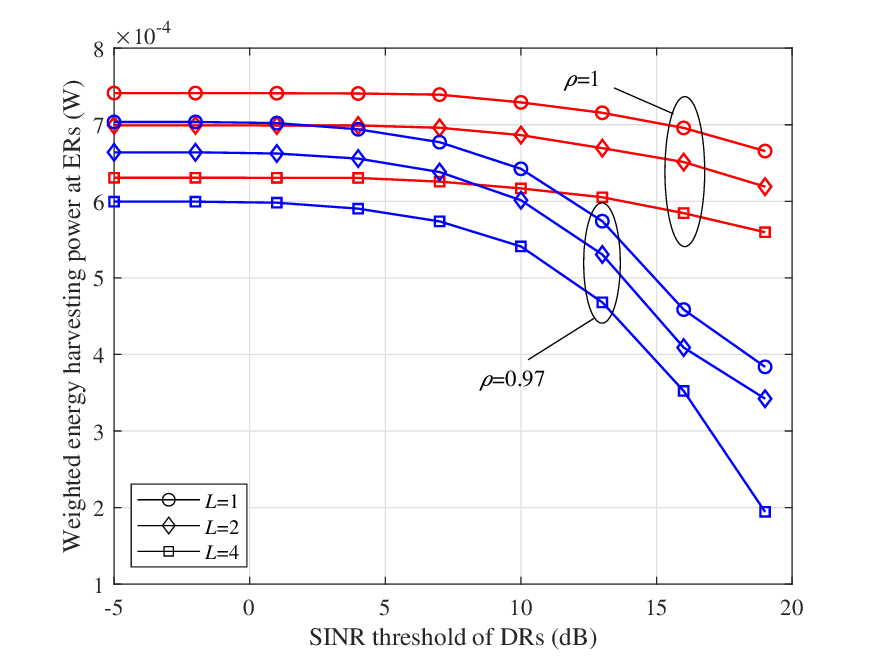}
	\caption{Weighted energy harvesting power at ERs versus the SINR threshold with different $L$ and $\rho$.}
	\label{fig:IDET_1}
\end{figure}

Fig. \ref{fig:IDET_1} depicts the weighted energy harvesting power at ERs versus the SINR threshold with different $L$ and $\rho$ in FA assisted WET system. The antenna size and port number are set as $W=0.5$ and $N=200$, respectively. Observe from Fig. \ref{fig:IDET_1} that when the channel estimation accuracy parameter $\rho$ is closer to 1, the DRs harvest more energy. Note that when $\rho=1$ and $L=1$, the CSI of all FA ports are estimated perfectly, which results in the best IDET performance. In addition, when $L$ becomes larger, there are fewer estimated port channels, resulting in the worse IDET performance. However, it is worthwhile to reduce the overhead of channel estimation by sacrificing the IDET performance, since it is impractical to acquire the CSI of all ports.
\section{Conclusion}\label{section:conclusion}
In this letter, we studied a FA assisted IDET system having multiple DRs and ERs, while the port selection and beamforming design were jointly optimized, in order to maximize the weighted energy harvesting power at ERs by guaranteeing the SINR at DRs. SDR and AO algorithms were then jointly applied to obtain the sub-optimal solutions. Numerical results showed that a larger port number and a larger size of fluid antenna always resulted in a better IDET performance. It was also evaluated that the performance of FA assisted IDET outperformed that of traditional MIMO counterpart having the same antenna size, which demonstrated that FA was more suitable for the low-power devices with limited hardware size.

\bibliography{Reference}
\end{document}